\documentclass[AMA,STIX1COL]{WileyNJD-v2}
\usepackage{moreverb}

\newcommand\BibTeX{{\rmfamily B\kern-.05em \textsc{i\kern-.025em b}\kern-.08em
T\kern-.1667em\lower.7ex\hbox{E}\kern-.125emX}}

\articletype{Research Article}%

\received{<day> <Month>, <year>}
\revised{<day> <Month>, <year>}
\accepted{<day> <Month>, <year>}

\usepackage{amsmath,amssymb,amsthm}
\usepackage{graphicx,psfrag,epsf}
\usepackage{enumerate}
\usepackage{url} 
\usepackage{multicol}
\usepackage{algorithm}
\usepackage{algpseudocode}
\usepackage{bbm}
\usepackage{xcolor}
\usepackage[all,cmtip]{xy}
\usepackage{pdfpages} 

\algnewcommand\algorithmicinput{\textbf{Input:}}
\algnewcommand\Input{\item[\algorithmicinput]}

\algnewcommand\algorithmiciterate{\textbf{Iterate}}
\algnewcommand\Iterate{\item[\algorithmiciterate]}

\algnewcommand\algorithmicdefine{\textbf{Define}}
\algnewcommand\Define{\item[\algorithmicdefine]}

\algnewcommand\algorithmiccompute{\textbf{Compute}}
\algnewcommand\Compute{\item[\algorithmiccompute]}

\algnewcommand\algorithmicinitialize{\textbf{Initialize}}
\algnewcommand\Initialize{\item[\algorithmicinitialize]}

\algnewcommand\algorithmicwhileloop{\textbf{While}}
\algnewcommand\Whileloop{\item[\algorithmicwhileloop]}

\algnewcommand\algorithmicendloop{\textbf{Return:}}
\algnewcommand\Endloop{\item[\algorithmicendloop]}

\newcommand{\indep}{\perp \!\!\! \perp}

\renewcommand{\thefootnote}{\alph{footnote}}

\newcommand\extrafootertext[1]{%
    \bgroup
    \renewcommand\thefootnote{\fnsymbol{footnote}}%
    \renewcommand\thempfootnote{\fnsymbol{mpfootnote}}%
    \footnotetext[0]{#1}%
    \egroup
}

\DeclareMathOperator{\EX}{\mathbb{E}}
\DeclareMathOperator*{\argmax}{arg\,max}
\DeclareMathOperator{\expit}{expit}

\theoremstyle{plain} 
\newtheorem{newlemma}{Lemma}

\begin{document}

\title{A comprehensive framework for the evaluation of individual treatment rules from observational data }

\author[1,2,3]{François Grolleau}

\author[1,2]{François Petit}

\author[1,2,3]{Raphaël Porcher}

\authormark{GROLLEAU \textsc{et al}}

\address[1]{\orgdiv{Université Paris Cité, Paris, France}}

\address[2]{\orgdiv{Centre de Recherche Épidémiologie et Statistiques (CRESS-UMR1153), INSERM, INRAE, Paris, France}}

\address[3]{\orgdiv{Centre d’Épidémiologie Clinique, Assistance Publique-Hôpitaux de Paris, Hôtel-Dieu, Paris, France}}

\corres{François Grolleau, Centre d’Épidémiologie Clinique, Hôpital Hôtel-Dieu, 1 Place du Parvis Notre-Dame Paris, 75004, Paris, France. \email{francois.grolleau@aphp.fr}
}
 

\abstract[Abstract]{Individualized treatment rules (ITRs) are deterministic decision rules that recommend treatments to individuals based on their characteristics. Though ubiquitous in medicine, ITRs are hardly ever evaluated in randomized controlled trials. To evaluate ITRs from observational data, we introduce a new probabilistic model and distinguish two situations: i) the situation of a newly developed ITR, where data are from a population where no patient implements the ITR, and ii) the situation of a partially implemented ITR, where data are from a population where the ITR is implemented in some unidentified patients. In the former situation, we propose a procedure to explore the impact of an ITR under various implementation schemes. In the latter situation, on top of the fundamental problem of causal inference, we need to handle an additional latent variable denoting implementation. To evaluate ITRs in this situation, we propose an estimation procedure that relies on an expectation-maximization algorithm. In Monte Carlo simulations our estimators appear unbiased with confidence intervals achieving nominal coverage. We illustrate our approach on the MIMIC-III database, focusing on ITRs for dialysis initiation in patients with acute kidney injury.}


\jnlcitation{\cname{%
\author{Grolleau F.}, 
\author{Petit F.}, 
\author{Porcher R.} (\cyear{2023}), 
\ctitle{A comprehensive framework for the evaluation of individual treatment rules from observational data},
\cjournal{arXiv:2207.06275}.}}

\maketitle

\extrafootertext{The authors gratefully acknowledge the Agence Nationale de la Recherche who partially funded this work under grant agreement no. ANR-18-CE36-0010-01. François Petit was supported by the IdEx Université Paris Cité, ANR-18-IDEX-0001. Raphaël Porcher acknowledges the support of the French Agence Nationale de la Recherche as part of the “Investissements d’avenir” program, reference ANR-19-P3IA-0001 (PRAIRIE 3IA Institute).}

\section{Introduction}
\label{sec:intro}
Individualized treatment rules (ITRs) are decision rules that recommend treatments to individuals based on their observed characteristics to maximize favorable outcomes on average. ITRs are widespread in medicine. In fact, most guidelines as well as the recently released computerized clinical decision support tools can be viewed as ITRs.\citep{Shalit2020, Sutton2020} Notable examples include decision tools for revascularization strategies in patients with coronary artery disease \citep{Takahashi2019} and for the personalization of blood pressure targets in hypertensive patients \citep{Basu2017}. For evaluating the impact of an ITR, the gold standard would be to conduct a randomized controlled trial (RCT) comparing the implementation of that ITR to usual care. Yet, there are practical challenges to conducting such RCTs. \citep{Tannock2016} As ITRs often recommend treatments similar to usual care, the expected population-level effect is likely small and necessitating very large sample sizes. Moreover, health agency oversight is less stringent for the implementation of ITRs than for drug compounds and so, both the incentives and funding opportunities for conducting RCTs of ITRs are scarce. In practice, these RCTs remain rare. As a result, many ITRs are being implemented despite the lack of evidence supporting their benefit. In this paper, we develop a framework to evaluate from observational data the impact of ITRs.\par 

To fit with most ITRs available in medicine (e.g., computerized clinical decision support tools, or guidelines), we view ITRs as deterministic maps recommending one of two treatment options. To make inference accounting for real-life prescription of treatment by physicians, we consider that deterministic ITRs are stochastically implemented with a probability of implementation depending on patient characteristics. Critically, we then distinguish two situations: i) the ITR was just released and treatment prescription was never based on it in the population, or ii) the ITR was available and, for some patients in the population, treatment prescription was based on it. We term these two situations the \emph{new ITR} and the \emph{partially implemented ITR} situations, respectively. In the former situation, we propose to numerically explore the benefit an ITR may have under different implementation schemes. In the latter situation, inference is more challenging as we are typically given observational data where we do not know which patients had implemented the ITR. That is, on top of the fundamental problem of causal inference, we need to handle an additional latent variable denoting implementation. To address this situation, we develop a new probability model and rely on a mixture of experts fitted via an EM algorithm for inference. \par

ITR estimation and evaluation has been considered in the literature of both statistics \citep{Qian2011,Zhao2012,Luedtke2016} and machine learning.\citep{Kallus2018,Thomas2016} Works most related to ours include the evaluation of stochastic rules,\citep{Diaz2013} biomarker performance,\citep{Janes2014} and ITR value accounting for the number of treated units. \citep{Imai2021} We note that encouragement designs \citep{Baiocchi2014} and instrumental variable \citep{Angrist1995,Frangakis2002} methods tackle problems related to, but subtly different from the question of estimating the effect of an ITR. To our knowledge, no work has focused on data originating from a \emph{partially implemented ITR situation}, nor pursued to develop a comprehensive framework for the evaluation of ITRs from observational data. \par

This article is organized as follows. In the evaluation metrics section, we introduce our causal model as well as our three estimands of interest: the Average Rule Effect (ARE), the Average Implementation Effect (AIE), and the Maximal Implementation Gain (MIG). In the inference section, we provide a method to estimate the ARE, AIE, and MIG and compute their standard error in both the \emph{new ITR} and \emph{partially implemented ITR} situations. In the simulation section, we study the properties of our estimators in the more challenging \emph{partially implemented ITR situation}. Finally, in the application section, we illustrate our approach on the MIMIC-III database, focusing on ITRs for dialysis initiation in patients with acute kidney injury. We evaluate two ITRs corresponding to the \emph{new ITR} and the \emph{partially implemented ITR} situations. The computer code for simulation studies and data applications is available at \url{https://github.com/fcgrolleau/ITReval}.

\section{Setup and evaluation metrics}

Following Neyman-Rubin causal model, we consider that a patient with observed outcome $Y$ has two potential outcomes $Y^{a=0} $ and $Y^{a=1}$ representing the outcome s/he would achieve if, possibly contrary to fact, s/he had received treatment option $A = 0$ or $A = 1$ respectively.\citep{Neyman1923,Rubin1974} Without loss of generality, we consider $A=1$ indicates that a patient received a specific treatment, and $A = 0$ indicates s/he received a control. Additionally, we consider for each patient, a vector of pre-treatment covariates $X$ with values in  $\mathcal{X}$. 

We assume that we are given an ITR that is, a deterministic map $r\colon \mathcal{X} \mapsto \{0;1\}$ which assigns a treatment option to each patient with covariates $x$. We model the implementation of the rule by the binary random variable $S$ where $S=1$ indicates that a patient's physician consulted the ITR and followed its recommendation to prescribe treatment (e.g., the physician accesses the ITR and takes its recommendation seriously). For the rest of this paper, we term $S=1$ as implementing the ITR. On the contrary, $S=0$ indicates that a patient's physician either did not consult the ITR or consulted it but prescribed a treatment opposite to what the ITR recommended (e.g., the physician has no access to the ITR or s/he does not take its recommendation seriously). The formal definition of $S=1$ and $S=0$ is given in Table \ref{table:Srecap}. Note that when $S=0$ the prescribed treatment may still match the ITR's recommended treatment—i.e., the physician did not consult the ITR but, based on other grounds, s/he prescribed a treatment identical to the ITR's recommendation. We define the stochastic implementation function as the conditional distribution $\rho(x)=\EX[S| X=x]$.\footnote{Note that we consider here the stochastic implementation of a deterministic rule $r$ through $\rho$. This is different from defining a function $\mathcal{X} \rightarrow [0; 1]$ which would assign to each value $x$ a probability to allocate treatment $A=1$. This would correspond to what we call a stochastic rule — which $r$ is not.} 

\begin{table}[h!]
\begin{center}
\caption{The precise definition of $S=1$ (ITR is implemented) and $S=0$ (ITR is not implemented).\label{table:Srecap}}

\begin{tabular}{ccclllclll}
\cline{3-10}
                                            &     & \multicolumn{8}{l}{A physician consults the ITR}   \\ \cline{3-10} 
                                            &     & \multicolumn{4}{c}{No}   & \multicolumn{4}{c}{Yes} \\ \hline
The prescription matches the ITR's & No  & \multicolumn{4}{c}{S=0}  & \multicolumn{4}{c}{S=0} \\
recommendation i.e., $A = r(X)$               & Yes & \multicolumn{4}{c}{S=0}  & \multicolumn{4}{c}{S=1} \\ \hline
\end{tabular}
\end{center}
\end{table}

We define the propensity score $\pi$ as the conditional distribution $\pi(x)=\EX[A| X=x]$ and the treatment-specific prognostic functions $\mu_0$, $\mu_1$ as the functions satisfying $\mu_1(x)=\EX[Y^{a=1}|X=x]$ and $\mu_0(x)=\EX[Y^{a=0}|X=x]$. We denote $\tau$ the conditional average treatment effect (CATE) function i.e., $\tau(x)=\EX[Y^{a=1}-Y^{a=0}|X=x]=\mu_1(x)-\mu_0(x)$. 

\subsection{A probability model for the data generating mechanism}

Our goal in this subsection is to introduce a new causal model that allows us to determine the causal effect of implementing versus not implementing an ITR. We introduce $A^{s=1}$, the potential treatment that would be given to a patient if her/his physician implemented the ITR, i.e., $A^{s=1} = r(X)$, and $A^{s=0}$ the potential treatment s/he would be given if her/his physician did not implement the ITR. Similarly, we define $Y^{s=1}$ and $Y^{s=0}$, patient's potential outcomes when physicians do or do not implement the ITR, respectively. In addition, we define the variables with superscript $(-)^{*}$ the counterfactual variables that are observable in the situation where some physicians sometimes implement the ITR. The variables $S^*,A^*,$ and $Y^*$ respectively indicate the implementation status, the prescribed treatment, and the outcome in this situation where the ITR is partially implemented. We consider that we can easily identify which of the following two situations we are dealing with:
\begin{description}
    \item[A.] A situation where physicians never implement the ITR because it was not available. In this situation, the variables with superscript $(-)^{s=0}$ are observed and we have $S=0, A=A^{s=0}, Y=Y^{s=0}$. From this point onward, we call this situation the \emph{new ITR situation}.
    \item[B.] A situation where some physicians sometimes implement the ITR to prescribe treatment. In this situation, the variables with superscript $(-)^*$ are all observable (though in practice $S^*$, the implementation status, is often not collected) and we have $S=S^{*}, A=A^{*}, Y=Y^{*}$. For the remainder of this paper, we refer to this situation as the \emph{partially implemented ITR situation}.
\end{description}
Importantly, we consider that all counterfactual variables exist in both situations. However, we consider that these two situations are mutually exclusive. In particular, we consider that we can straightforwardly identify which of these two situations we are dealing with. That is, formally, we consider we can always observe a binary random variable $U$ where $U = 1$ indicates we are in a \emph{partially implemented ITR situation}, and $U = 0$ indicates we are in a \emph{new ITR situation}. This variable $U$ allows us to precisely formulate the link between observed random variables and potential outcomes as
\begin{align*}
    Y &=U Y^{*} + (1-U) Y^{s=0}, \\
    S &= U S^{*}, \\
    A &= U A^{*} + (1-U) A^{s=0}.
\end{align*}
 In this work, we do not consider counterfactuals with respect to $U$. Rather, to evaluate the effect of an ITR, we focus on the variables $Y^{s=0},Y^{s=1},$ and $Y^*$. For clarity, in the remainder of this paper, we avoid making reference to $U$. To identify causal effects, we rely on the subsequent assumptions of exclusion restriction, exchangeability, and overlap. 

\begin{assumption}[{\bf exclusion restriction}]\label{as1} The effect of the ITR on the outcome $Y$ is only mediated through the treatment, that is,
\begin{align}
Y^{s=1}=A^{s=1} Y^{a=1}+ \big(1-A^{s=1}\big) Y^{a=0}, \label{eq:consitency_ys1} \\
Y^{s=0}= A^{s=0}Y^{a=1} + (1-A^{s=0})Y^{a=0}, \label{eq:consitency_ys0} \\
A^\ast=S^\ast A^{s=1}+(1-S^\ast)A^{s=0}, \label{eq:consitency_ttt} \\
Y^\ast=A^\ast Y^{a=1}+ (1-A^\ast) Y^{a=0}. \label{eq:consitency_ITR}
\end{align}
\end{assumption}

\begin{assumption}[{\bf exchangeability}]\label{as2} All confounders and variables causing implementation are measured, that is,
\begin{align} 
\{Y^{a=1},Y^{a=0}\}   \indep A^{s=0}|X, \label{eq:ex_as0} \\
\{Y^{a=1},Y^{a=0}\}   \indep A^{*}|X,  \label{eq:ex_a*} \\
A^{s=0}   \indep S^{*}|X \label{eq:ex_S*} .
\end{align}
\end{assumption}

\noindent In the \emph{new ITR situation}: 
\begin{itemize}
\itemsep-.5em 
    \item[-] equation \eqref{eq:ex_as0} has the same interpretation as the classic no unmeasured confounder assumption,
    \item[-]  the interpretation of equation \eqref{eq:ex_a*} is that we already record all the confounding variables that will be in effect under the future implementation of the ITR,\footnote{Assuming that we measured all the prognostic variables and that these will not have changed in the future is enough to satisfy this assumption.}
    \item[-] the interpretation of equation \eqref{eq:ex_S*} is that we have measured the variables that may cause future implementation of the ITR.
\end{itemize}
In the \emph{partially implemented ITR situation}:
\begin{itemize}
\itemsep-.5em 
    \item[-] equation \eqref{eq:ex_a*} has the same interpretation as the classic no unmeasured confounder assumption,
    \item[-] the interpretation of equation \eqref{eq:ex_as0} is that, for the patients whose physician did not implement the ITR, all the confounding variables are measured,
    \item[-] the interpretation of equation \eqref{eq:ex_S*} is that we have measured the variables that cause the current implementation of the ITR.
\end{itemize}

\begin{assumption}[{\bf overlap}]\label{as3} Within all realistic levels of covariates, the patients could receive either treatment—including in the absence of ITR implementation. That is, denoting $\pi^{s=0}$ the propensity score in the absence of ITR implementation, i.e., $\pi^{s=0}(x)=\EX[A^{s=0}|X=x],$
\begin{align*}
 \forall \ x \in \mathcal{X}, \quad 0 < \pi(x)   < 1, \quad \text{and} \quad   0 < \pi^{s=0}(x)  < 1.
\end{align*} 
\end{assumption}

Observing the overlap assumption, we see that as the propensity score functions $\pi$ and $\pi^{s=0}$ can never be deterministic rules, they are thus stochastic rules. We define two additional stochastic rules: the propensity score under stochastic implementation $\pi^\ast$ that is $\pi^\ast(x)=\EX[A^*| X=x]$, and the stochastic implementation function under implementation $\rho^\ast$ i.e., $\rho^\ast(x)=\EX[S^*| X=x]$.

\begin{figure}[htb!]
\begin{equation*}
\xymatrix{
            \mathbf{A.}&&&& \mathbf{B.} &&      &   X\ar[dll]\ar[d]\ar[drr]\ar[ddll] \ar[ddrr]   &       &         \\
    &&&&& A^{s=0}\ar[drr]\ar@/_1.3pc/[dd] &       &   S^*\ar[d]   &       & A^{s=1}\ar[dll]\ar@/^1.3pc/[dd]  \\
     X \ar[r]\ar@/^2pc/[rr] &  A^{s=0}\ar[r] & Y^{s=0} &&& Y^{a=0}\ar[d]\ar[ddrr]\ar[drrrr] &       &   A^*\ar[dd]   &       & Y^{a=1} \ar[d]\ar[dllll]\ar[ddll]  \\
    &&&&&Y^{s=0} &       &             &       & Y^{s=1}  \\
            &&&&&&    &       Y^*      &      &         
    } 
\end{equation*}
\caption{
The probabilistic graphical model associated with the data generating mechanism in the \emph{new ITR situation} (Panel A) and the \emph{partially implemented ITR situation} (Panel B).}
\label{fig:graph_dgp}
\end{figure}
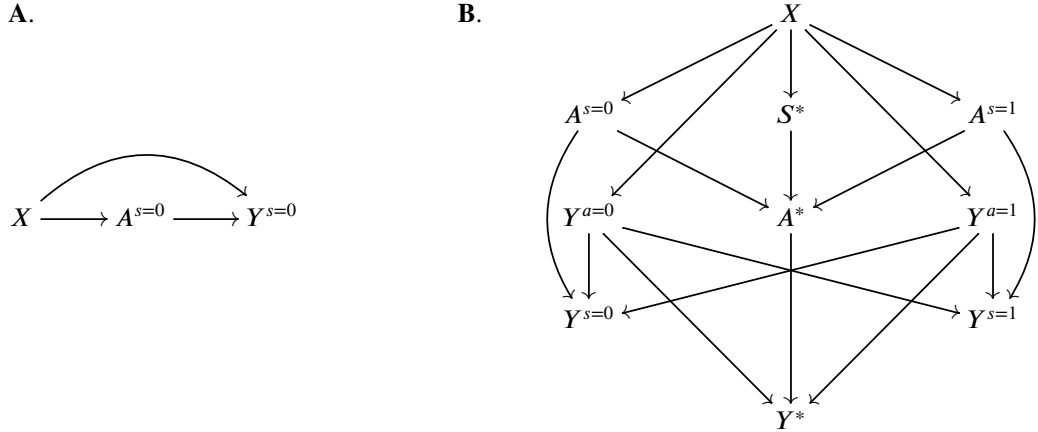

Summarizing the exclusion restriction equations \eqref{eq:consitency_ys1}, \eqref{eq:consitency_ys0}, \eqref{eq:consitency_ttt}, \eqref{eq:consitency_ITR} and the exchangeability equations \eqref{eq:ex_as0},  \eqref{eq:ex_a*}, \eqref{eq:ex_S*}, the data generating mechanism in the \emph{new ITR} and the \emph{partially implemented ITR} situations can be represented by the probabilistic graphical models in Figure \ref{fig:graph_dgp}A and \ref{fig:graph_dgp}B, respectively. 
\\

In our setup, it will prove convenient to define $q_1$, the prognostic function under ITR implementation, as $q_1(x)=\EX[Y^{s=1}|X=x]$ and $q_0$, the prognostic function in the absence of ITR implementation, as $q_0(x)=\EX[Y^{s=0}|X=x]$. Conditioning equation \eqref{eq:consitency_ys1} with respect to $X$ leads to $q_1(x)=r(x)\mu_1(x)+\{1-r(x)\}\mu_0(x)$.

\subsection{Estimands of interest}

We now introduce three estimands. First, the \textit{Average Rule Effect} (ARE) of an ITR $r$:
\begin{equation*}
     \Delta(r) = \EX[Y^{s=1} - Y^{s=0}].
\end{equation*}

This represents the population-level effect of the ITR on outcome $Y$ in a randomized trial comparing a group where patients are systematially given the treatment recommended by ITR to usual care in the absence of ITR implementation.


Second, we define the \textit{Average Implementation Effect} (AIE) of $r$ as 
\begin{equation*}
     \Lambda(r,\rho^\ast) = \EX[Y^\ast - Y^{s=0}].
\end{equation*}

This represents the population-level effect of the ITR on outcome $Y$ in a randomized trial comparing a group where physicians are provided with the ITR's treatment recommendation to usual care under no implementation. We may thus consider that in the experimental group the treatment $A$ is prescribed according to a stochastic implementation $\rho^\ast$ of $r$. Note that the ARE can be considered as a special case of the AIE where the stochastic implementation is perfect ($\rho^\ast \equiv 1$). We however single it out because this representation allows to assess whether the ITR has a potential population-level benefit, or if it is instead poorly designed.

Last, we define the \textit{Maximal Implementation Gain} (MIG) of $r$:
\begin{equation*}
     \Gamma(r,\rho^\ast) = \EX[Y^{s=1} - Y^\ast].
\end{equation*}
This represents the difference in average outcome between a full implementation of the ITR and the current or future partial implementation of the rule. From these definitions, it follows that 
\begin{equation*}
\Delta(r) = \Lambda(r,\rho^\ast) + \Gamma(r,\rho^\ast). 
\end{equation*}

\section{Inference}

We assume we are given a single random sample of $n$ independent and identically distributed (i.i.d.) units $(X^T_i, A_i, Y_i)_{1\leq i \leq n}$ from a target population. As in the previous section, we distinguish between data originating from \emph{new} and \emph{partially implemented} ITR situations:

\begin{description}
\item[A.] In data originating from the \emph{new ITR situation}, we have $S_i = 0$, $A_i=A_i^{s=0}$, and $Y_i=Y_i^{s=0}$. Note that the second equality implies $\pi = \pi^{s=0}$. In this situation, we drop all $(-)^{s=0}$ superscripts and use $\pi$ rather than $\pi^{s=0}$ for clarity. Clearly, estimation of the AIE and MIG is not possible from data alone in this situation. Nonetheless, in the next section, we propose to explore their behaviour by hypothesizing implementation schemes.  
\item[B.] In data originating from the \emph{partially implemented ITR situation}, we have $S_i=S_i^\ast$, $A_i=A_i^\ast$, and $Y_i=Y_i^\ast$. Note that the first two equalities imply  $\rho = \rho^*$ and $\pi = \pi^*$. In this situation, we thus drop all $(-)^\ast$ superscripts for clarity. Because we expect that $S_i$ will not have been collected in this situation, we treat it as a latent variable.
\end{description}

\subsection{New individualized treatment rule situation}\label{subsub:newITR}
\subsubsection{Average rule effect}\label{subsub:newimpARE}
Using exclusion restriction \eqref{as1}, exchangeability \eqref{as2}, and positivity \eqref{as3}, in the \emph{new ITR situation}, we have
\begin{align*}
    \Delta(r) &= \EX\big[q_1(X)-Y\big]  \\
    &=\EX\Bigg[ \Big\{r(X)\frac{A}{\pi(X)}+\{1-r(X)\}\frac{1-A}{1-\pi(X)}-1 \Big\}Y \Bigg].
\end{align*}
These equations suggest the following two estimators for $\Delta(r)$
\begin{equation}\label{est:are_q_new}
    \widehat{\Delta}_{Q}(r) = n^{-1} \sum_{i=1}^{n} r(X_i)\hat{\mu}_1(X_i)+\{1-r(X_i)\}\hat{\mu}_0(X_i)-Y_i,
\end{equation}

\begin{equation}\label{est:are_ipw_new}
    \widehat{\Delta}_{IPW}(r) = n^{-1} \sum_{i=1}^{n} \Big[r(X_i) \frac{A_i}{\hat{\pi}(X_i)}+\{1-r(X_i)\}\frac{1-A_i}{1-\hat{\pi}(X_i)}-1 \Big]Y_i,
\end{equation}
where as for all estimators proposed hereafter, $\mu_0(\cdot)$, $\mu_1(\cdot)$, and $\pi(\cdot)$ can be estimated via any supervised learning method from observations $(X_i^\mu, Y_i)_{i:A_i=0}$, $(X_i^\mu, Y_i)_{i:A_i=1}$, and $(X_i^\pi, A_i)_{1 \leq i \leq n}$ respectively.\footnote{Here, $X_i^{\mu}$ and $X_i^{\pi}$ denote two subsets of the relevant variables contained in $X_i$.}
An augmented counterpart of these estimators can be derived from Zhang et al.\citep{Zhang2012}:
\begin{align}\label{est:are_aipw_new}
    \widehat{\Delta}_{AIPW}(r) = n^{-1} \sum_{i=1}^{n} \Big[ \frac{\mathcal{C}_i^r Y_i}{\hat{\pi}(X_i)\mathcal{C}_i^r + \{1-\hat{\pi}(X_i) \}(1-\mathcal{C}_i^r)}
    -\frac{\mathcal{C}_i^r - [\hat{\pi}(X_i)\mathcal{C}_i^r + \{1-\hat{\pi}(X_i)  \}(1-\mathcal{C}_i^r) ] }{\hat{\pi}(X_i)\mathcal{C}_i^r + \{1-\hat{\pi}(X_i) \}(1-\mathcal{C}_i^r)}\hat{q}_1(X_i) - Y_i \Big] 
\end{align}
where we set $\mathcal{C}_i^r =\mathbbm{1}\{r(X_i)=A_i\}$ and $\hat{q}_1(X_i)=r(X_i)\hat{\mu}_1(X_i)+\big\{1-r(X_i)\big\}\hat{\mu}_0(X_i)$ for clarity. We refer the reader to Tsiatis et al. \citep[section 3.3.3 p. 61]{Tsiatis2019} for an extensive study of this specific case and the derivation of approximate large sample distribution.
Using the CATE, the ARE can also be reformulated as
\begin{equation}\label{eq:are_ite_new}
     \Delta(r) = \EX\Big[\{r(X) -\pi(X)\}\tau(X)\Big]
\end{equation}
(a proof is given in Appendix A). This leads to the following estimator
\begin{equation*}
    \widehat{\Delta}_{CATE}(r) = n^{-1} \sum_{i=1}^{n} \{r(X_i) -\hat{\pi}(X_i)\}\hat{\tau}(X_i).
\end{equation*}
Though the latter estimator requires to estimate the CATE $\tau$, and hence may be less practical than estimators \eqref{est:are_q_new}, \eqref{est:are_ipw_new}, or \eqref{est:are_aipw_new}, the equation \eqref{eq:are_ite_new} makes explicit the respective contribution of $\tau$, $r$ and $\pi$ to the ARE.

\subsubsection{AIE, MIG under the modeling of the stochastic implementation functions}\label{sif_models}

When the ITR is new and has never been deployed, the way in which it will be implemented is unpredictable. Hence, the AIE and the MIG cannot be estimated from data alone. However, it can be interesting to study numerically how the AIE and MIG would vary under different stochastic implementation schemes as this can provide information about the appropriateness of future ITR deployment. In the \emph{new ITR situation}, it is possible to show that 
\begin{align}
    \Lambda(r, \rho^\ast) 
    &= \EX\Big[\{\pi^\ast(X)-\pi(X)\}\tau(X)\Big] \label{eq:aie_pi_ite_new} \\
    &= \EX\Big[\rho^\ast(X)\{r(X)-\pi(X)\}\tau(X)\Big], \label{eq:aie_ite_new}
\end{align}
and
\begin{align*}
    \Gamma(r, \rho^\ast) 
    &= \EX\Big[\{r(X)-\pi^\ast(X)\}\tau(X)\Big]\\
    &= \EX\Big[\{1-\rho^\ast(X)\}\{r(X)-\pi(X)\}\tau(X)\Big]
\end{align*}
(a proof is given in Appendix B). Hence, given an estimate $\hat{\tau}$ of the CATE function \citep[see Jacob][for a review of the available estimation methods]{Jacob2021}, an estimate $\hat{\pi}$ of the propensity score and a numerical model of $\rho^\ast$ for a future stochastic implementation function, estimates of the AIE and the MIG are computable via

\begin{equation*}
    \widehat{\Lambda}_{CATE}(r, \rho^\ast) = n^{-1} \sum_{i=1}^{n} \rho^\ast(X_i)\{r(X_i) -\hat{\pi}(X_i)\}\hat{\tau}(X_i),
\end{equation*}
and
\begin{equation*}
    \widehat{\Gamma}_{CATE}(r,\rho^\ast) = n^{-1} \sum_{i=1}^{n} \{1-\rho^\ast(X_i)\}\{r(X_i) -\hat{\pi}(X_i)\}\hat{\tau}(X_i).
\end{equation*}

Below, we propose, three schemes that model the form the stochastic implementation function may take in future deployment of the ITR:

\begin{description}
	\item[$\bullet$] The random implementation scheme, where we model $\rho^\ast(\cdot)$ as
	\begin{equation*} \label{rd-eq}
	\rho^\ast_{rd,\alpha}(x) = \alpha
	\end{equation*}
with $\alpha \in [0;1]$ a parameter modelling the random implementation such that, uniformly for all patients, higher values of $\alpha$ are associated with higher probabilities of following the rule. This model of $\rho^\ast$ describes a situation where patients are treated according to an implementation of the rule at random with probability $\alpha$ regardless of their characteristics.
	\item[$\bullet$] The cognitive bias scheme, where we model $\rho^\ast(\cdot)$ as
	\begin{equation*} \label{cb-eq}
	\rho^\ast_{cb,\alpha}(x) = \{1-\left | r(x) - \pi(x) \right |\}^{ \frac{1}{2} \log \frac{\alpha+1}{1-\alpha}} 
	\end{equation*}
with $\alpha \in [0;1[$ a cognitive bias parameter such that higher values of $\alpha$ are associated with lower probabilities of following the rule for a given gap between the recommendation from the ITR and usual care under no implementation. This implementation scheme describes a situation where physicians follow the ITR recommendation more often when recommendations are similar to current practices and this trend to resist change increases as $\alpha$ increases.
	\item[$\bullet$] The confidence level scheme, where we assume that the ITR was constructed from estimated CATEs, $\tilde{\tau}(x)$, as in for instance $r(x)= \mathbbm{1} [\tilde{\tau}(x)<0 ],$ when $Y=1$ denotes mortality.  For this scheme to be actionable, $\tilde{\tau}(x)$ and their standard errors $se_{  \tilde{\tau}(x) }$ must be provided along the ITR they helped build. Under such conditions, we model $\rho^\ast(\cdot)$ as
	\begin{equation*} \label{ci-eq-abs}
	\rho^\ast_{cl,\alpha}(x) = \mathbbm{1}  [{ \{ \tilde{\tau}(x) - q_ { 1 - \alpha /2 } se_{  \tilde{\tau}(x) } \}  \{ \tilde{\tau}(x) + q_ { 1 - \alpha /2 } se_{ \tilde{\tau}(x) } \}   > 0 }]
	\end{equation*} 
	for CATEs  provided on an absolute scale (i.e., individual absolute risk difference) with $\alpha \in [0;1]$ a type I error parameter such that smaller values of $\alpha$ lead to wider confidence intervals for $\tilde{\tau}(x)$. This scheme describes a situation where physicians follow the ITR recommendation only when there is evidence that $\tau(x) \neq 0$ at significance level $\alpha$.
\end{description}

\subsubsection{Illustrative examples}
In this section, we aim to provide a sense of what our method is trying to achieve when applied in the \emph{new ITR situation}. For that purpose, in this subsection, we provide a toy model. Observing Formula \eqref{eq:aie_pi_ite_new}, we see that the AIE of a new ITR $r$ gets far off from zero as the difference between current treatment allocation and future treatment allocation under a stochastic implementation of $r$ increases. More precisely, observing Formula \eqref{eq:aie_ite_new}, we note that the AIE of a new ITR $r$ gets far off from zero as patients with common levels of covariates $x$ have i)
a high probability $\rho^\ast(x)$ of implementing the rule,
and/or ii)
a difference $r(x)-\pi(x)$ between recommendation from the ITR and usual care under no implementation far off from zero,
and/or iii) large CATEs $\tau(x)$.

\begin{figure}[htb!]
    \centering
    \includegraphics[width=0.775\textwidth]{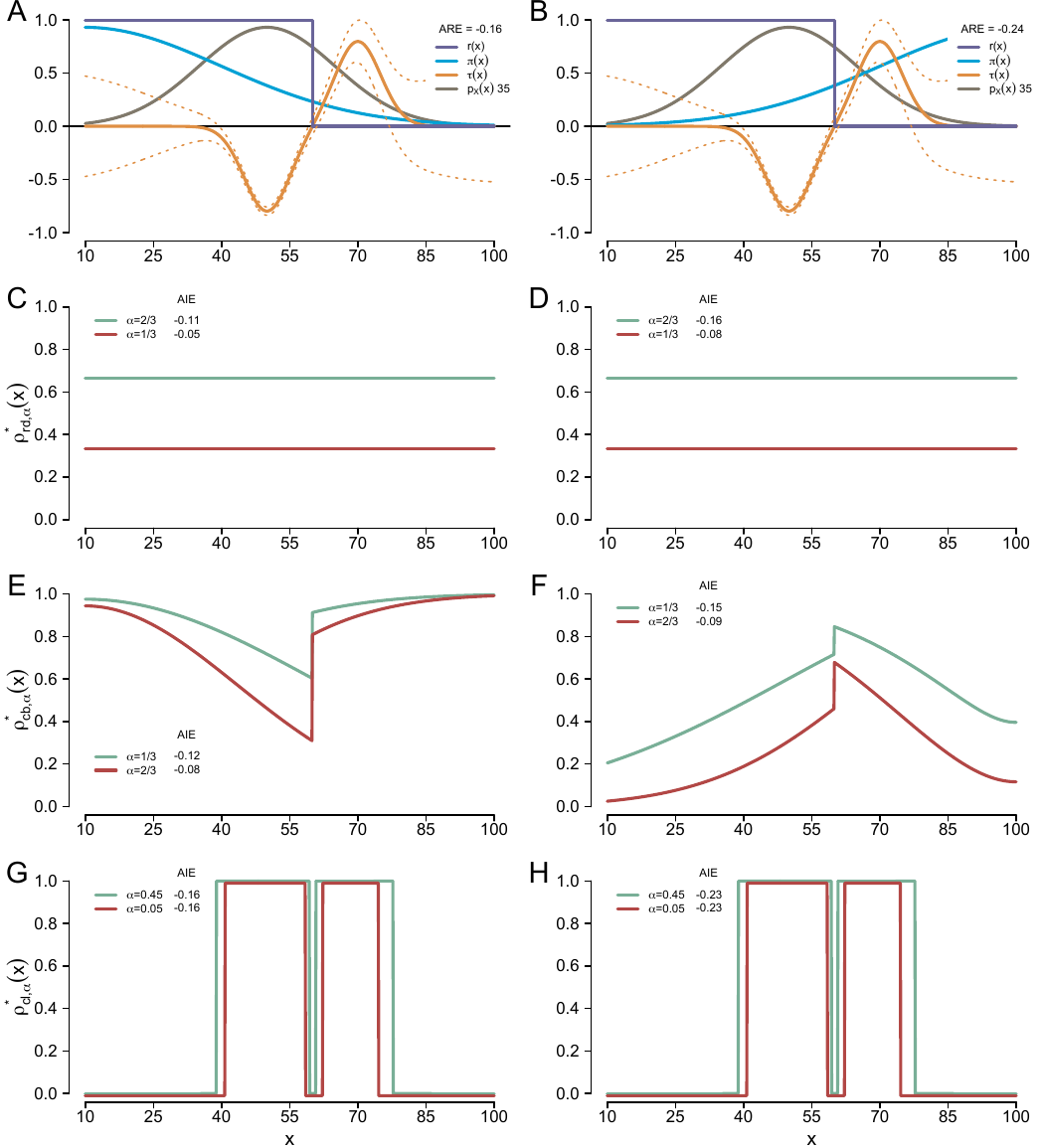}
    \caption{ Panel A displays our first illustrative example where under no implementation usual care is such that younger patients are treated more often. 
    Panel B displays our second illustrative example where under no implementation usual care is such that older patients are treated more often. The random implementation scheme is given panels C and D for both examples, respectively, the cognitive bias scheme on panels E and F, and the confidence level scheme panels G and H.
    The dotted lines correspond to the CATEs plus/minus its standard errors. AIEs are reported for each of implementation schemes and lower values indicate greater benefit from ITR implementation. We denote $p_X$ the probability density of $X$ (we re-scale $p_X(x)$ by a factor $35$ for illustration purposes).  
    }
    \label{fig:ex1}
\end{figure}

For illustration purposes, we imagined a disease for which only one patient characteristic, the age $x$, is relevant to treatment decision-making. In a population of patients with mean age 50 (standard deviation 15), we wish to evaluate the effectiveness of an ITR $r$ with respect to the occurrence of an unfavorable binary outcome (i.e., 10-year mortality). 
In our two examples, ground truth is such that the treatment is beneficial for patients aged 40 to 60, detrimental for patients aged 60 to 80, and has almost no effect outside these ranges. For the sake of simplicity, we suppose that in both examples $r$ is $r(X) = \mathbbm{1} [{\tau(X)<0}]$ that is, the rule is optimal (Figure \ref{fig:ex1} Panels A and B).

In our first example (Figure \ref{fig:ex1}A), the usual care under no implementation is such that younger patients are treated more often while in our second example (Figure \ref{fig:ex1}B), older patients are treated more often. 
In the random implementation schemes (Figure \ref{fig:ex1} Panels C and D), physicians follow the ITR's recommendation at random with probability $1/3$ (red lines) or $2/3$ (green lines). In the cognitive bias schemes (Figure \ref{fig:ex1} Panels E and F), physicians follow the ITR's recommendation more often when the ITR's recommendation tracks the usual care under no implementation. Cognitive bias parameter is $2/3$ (red lines) or $1/3$ (green lines), and higher parameter values are associated with lower probabilities of complying with the ITR. In the confidence level schemes (Figure \ref{fig:ex1} Panels G and H), physicians follow the ITR's recommendation only when confidence intervals for the predicted CATEs do no cross zero. Type I error parameters for the confidence intervals are $0.05$ (red lines) or $0.45$ (green lines) with higher values associated with tighter confidence intervals and therefore higher probabilities of implementing the ITR.

Despite the fact that both examples relied on the implementation of an identical ITR based on the true CATE function, the population-level benefit of this ITR is different between examples for all schemes. The ARE of the deterministic rule was $-0.16$ in the population from example 1 and $-0.24$ in the population from example 2, indicating an $8\%$ greater benefit of implementing the ITR in population 2 than in population 1 if physicians always followed the ITR's recommendation. Similarly, in the stochastic implementation schemes, the population-level benefit of the ITR differ in the two example populations. In the random implementation scheme, AIEs are $-0.11$ and $-0.05$ in population 1 (Figure \ref{fig:ex1}C) versus $-0.16$ and $-0.08$ in population 2 (Figure \ref{fig:ex1}D) for random implementation parameters $2/3$ and $1/3$ respectively. We find similar differences in AIEs between population 1 and population 2 in the cognitive bias scheme (Figure \ref{fig:ex1} Panels E and F) and confidence level scheme (Figure \ref{fig:ex1} Panels G and H).

\subsection{Partially implemented individualized treatment rule situation} \label{subsub:iplITR}

In this subsection, our aim is to estimate the ARE, AIE and MIG with data sampled from a population where the ITR $r$ is partially implemented. Two cases have to be distinguished depending on whether the variable $S$ is collected. If $S$ is collected, estimating the ARE, AIE and MIG can be achieved by using a suitable adaptation of the IPW/AIPW estimators for the average treatment effect.\citep{Lunceford2004} However, in practice, we expect that the variable $S$ will not have been collected. Hence, we focus our attention on the case where $S$ needs to be regarded as a latent variable. Recall that in this subsection, we are dealing with a partially implemented ITR where the observed treatment follows either from the ITR being implemented or from the physicians disregarding the ITR to make treatment decisions. Inference in the partially implemented ITR situation relies on assumptions (\ref{as1}-\ref{as3}). Because in this situation estimation of the MIG is more straightforward than estimation of the ARE and AIE, we distinguish between the two cases.

\subsubsection{Maximal implementation gain}\label{subsub:impMIG}
We start by studying the MIG, as neither $S$ nor $\rho$ play a role for this estimand in the \emph{partially implemented ITR situation}. In fact in this situation, using exclusion restriction \eqref{as1}, exchangeability \eqref{as2}, and positivity \eqref{as3}, we have
\begin{align*}
    \Gamma(r,\rho)&= \EX\big[q_1(X)-Y\big] \\
     &= \EX \Bigg[ \Big\{r(X) \frac{A}{\pi(X)}+\{1-r(X)\}\frac{1-A}{1-\pi(X)}-1 \Big\}Y \Bigg].
\end{align*}
This suggests the estimators
\begin{equation*}
\widehat{\Gamma}_{Q}(r,\rho)=n^{-1} \sum_{i=1}^n   r(X_i)\hat{\mu}_1(X_i)+\{1-r(X_i)\}\hat{\mu}_0(X_i)-Y_i,
\end{equation*}
and
\begin{equation*}
  \widehat{\Gamma}_{IPW}(r,\rho)= n^{-1} \sum_{i=1}^n  \Big[r(X_i) \frac{A_i}{\hat{\pi}(X_i)}+\{1-r(X_i)\}\frac{1-A_i}{1-\hat{\pi}(X_i)}-1 \Big]Y_i.
\end{equation*}
The derivation is similar to that of the ARE in the \emph{new ITR situation} (equations \ref{est:are_q_new} and \ref{est:are_ipw_new}). We refer the reader to section \ref{subsub:newimpARE}, equation \eqref{est:are_aipw_new} for an augmented version of this estimator.

\subsubsection{Average rule effect and average implementation effect}\label{subsec:em}

Note that the MIG estimand is distinct from the ARE and AIE in that it does not involve the expectation term $\EX(Y^{s=0})$. In contrast, the ARE and AIE depend on the pairs of expectations $\EX(Y^{s=1})$, $\EX(Y^{s=0})$ and $\EX(Y)$, $\EX(Y^{s=0})$ respectively. The quantity $\EX(Y)$ is straightforward to estimate. The expectation $\EX(Y^{s=1})$ can be estimated by various means, for instance by taking the expectation of $\hat{q}_1(X)$ as in section \ref{subsub:impMIG}.

\begin{figure}[htb!]
    \begin{equation*}
\xymatrix{
& & & & & \\
& & & & *+[o][F.]{A} \ar[u]_{\pi(X)} & \\
& & *+[F]{\substack{\text{Gating} \\ \text{Network}}} \ar@{}[l]|-<<<{\rho(\cdot)} \ar@{}@<1ex>[rrr]^(.098){}="a"^(.54){}="b" \ar@{|}^<<<<<{g_0} "a";"b"
\ar@{}@<-1ex>[rrr]^(.098){}="c"^(.85){}="d" \ar@{|}_<<<<<{g_1} "c";"d"
& & & \\
& & \ar[u]^<<<<{X^{\rho}} & *+[F]{\substack{\text{Expert} \\ \text{Network}}} \ar[uur]^<<{p_0} \ar@{}[l]|-<<<{\pi^{s=0}(\cdot)\quad} & & *+[F]{\substack{\text{Expert} \\ \text{Network}}} \ar[uul]_<<{r(X)} \ar@{}[l]|-<<<{r(\cdot)} &  \\
& & & \ar[u]_<<<<{X^{\pi^{s=0}}} & & \ar[u]_<<<<{X} }
    \end{equation*}
\caption{ The graphical representation of the mixture of experts fitted by Algorithm \ref{algo:1}. Note that we consider $r$ as a known deterministic expert network, while both the expert network $\pi^{s=0}$ and the gating network $\rho$ are unknown stochastic rules.}
\label{fig:algo}
\end{figure}
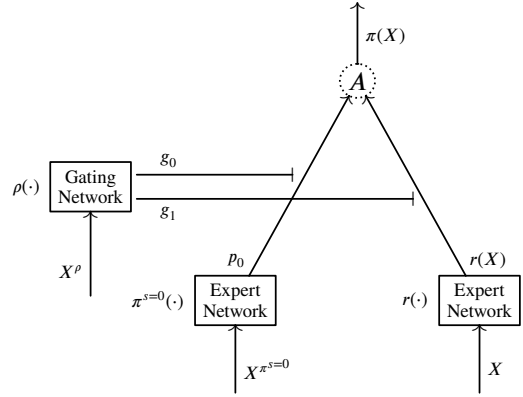

Estimation of $\EX(Y^{s=0})$ is more challenging than that of $\EX(Y^{s=1})$ because, substitution of $Y^{s=0}$ by its definition in \eqref{eq:consitency_ys0} involves the potential outcome $A^{s=0}$ which is not directly identifiable from equation \eqref{eq:consitency_ttt} as $S$ is a latent variable. Our approach to estimate $\EX(Y^{s=0})$ relies on the following result.
\begin{newlemma}
\label{lm1}
In the partially implemented ITR situation, the following relations holds
\begin{align}
\text{(i)} &\quad \pi(x) = \rho(x)r(x) +\{1-\rho(x)\}\pi^{s=0}(x), \label{model:score_prop} \\
\text{(ii)} &\quad q_0(x) = \pi^{s=0}(x)\mu_1(x) + \big\{1-\pi^{s=0}(x)\big\}\mu_0(x). \label{model:Q0}
\end{align}
\end{newlemma}
\noindent A proof of the lemma is given in Appendix C. Lemma \ref{lm1} suggests that the functions $\pi$ and $q_0$ may be represented by a particular type of mixture model called the mixture of experts model.\citep{Jordan1994} This model lays out a mixture of regression models (experts) where the proportions of the mixture (gating network) depend on the covariates. Observing equation \eqref{model:score_prop}, we see that $\pi$ can be viewed as a mixture of the known expert $r$ and the unknown expert $\pi^{s=0}$, while the proportions of the mixture are given by the unknown gating network $\rho$. Several authors, including Teicher;\cite{Teicher1963} Jiang and Tanner;\cite{Jiang1999} Allman et al.\cite{Allman2009} have studied the identifiability of mixture models. However, to the best of our knowledge, their results are not directly applicable to the specific type of mixture we are considering in equation \eqref{model:score_prop}. In Appendix D, we prove, under mild technical conditions on the treatment rule $r$, that if the covariables are continuous then the mixtures of experts considered in this paper are identifiable.
Equation \eqref{model:Q0} suggest to rewrite $\Delta(r)$ and $\Lambda(r,\rho)$ as 
\begin{align*}
  \Delta(r) &= \EX\big[q_1(X) - q_0(X)\big] \\
            &= \EX\Big[\{r(X)-{\pi^{s=0}}(X) \}\tau(X)\Big]\\
&\text{and}\\
  \Lambda(r, \rho) &= \EX\big[Y - q_0(X)\big] \\
            &= \EX\Big[Y - \mu_1(X)\pi^{s=0}(X) - \mu_0(X)\{1-\pi^{s=0}(X)\}\Big].
\end{align*}
Because $\pi^{s=0}$ is  unknown, we propose to estimate it via the procedure detailed in Algorithm \ref{algo:1}. This procedure details an EM algorithm, based on the fitting algorithm of Xu and Jordan\citep{Xu1993} and Jordan and Jacobs\citep{Jordan1994} where we posit parametric models for $\pi^{s=0}(\cdot)$ and $\rho(\cdot)$. In Appendix E, we detail how this algorithm was derived and how it could readily be extended to the case of non-parametric models. Figure \ref{fig:algo} depicts the graphical model for the approach to estimating $\pi^{s=0}(\cdot)$. Estimating $\mu_0(\cdot)$, $\mu_1(\cdot)$, and $\tau(\cdot)$ as in section \ref{subsub:newITR}, mixture of experts estimators for the ARE and AIE are then given by
\begin{align*}
    \widehat{\Delta}_{ME}(r)&=
   n^{-1} \sum_{i=1}^n \{r(X_i)-\hat{\pi}^{s=0}(X_i) \}\hat{\tau}(X_i),
\end{align*}
and
\begin{align*}
    \widehat{\Lambda}_{ME}(r,\rho)&= 
   n^{-1} \sum_{i=1}^{n}Y_i - \hat{\mu}_1(X_i)\hat{\pi}^{s=0}(X_i) - \hat{\mu}_0(X_i)\{1-\hat{\pi}^{s=0}(X_i)\}. 
\end{align*}
Assuming that all relevant parametric models are correctly specified, we note that each estimator $\widehat{\Delta}_{ME}(r)$ and $\widehat{\Lambda}_{ME}(r,\rho)$ jointly solve a set of “stacked” estimating equations. In particular, the parameters of $\pi^{s=0}(\cdot)$ jointly solve the derivative of the log-likelihood given in equation (E2) from the Appendix. The $\widehat{\Delta}_{ME}(r)$ and $\widehat{\Lambda}_{ME}(r,\rho)$ estimators can thus be viewed as M-estimators and it follows that under correct model specification, they are $\sqrt{n}$-consistent and asymptotically normal with asymptotic variances calculable via the empirical sandwich estimator. For a clear review of M-estimation and the stacked estimating equation method, we refer the reader to Stefanski and Boos.\citep{Stefanski2002} For simplicity, in the remainder of this paper, we propose to estimate the variance of the estimators $\widehat{\Delta}_{ME}(r)$ and $\widehat{\Lambda}_{ME}(r,\rho)$ via the boostrap. We assess the validity of this strategy in Monte Carlo simulations. 
\clearpage

\begin{algorithm}
\caption{ \scriptsize{The EM procedure for estimating $\pi^{s=0}(\cdot)$ in the partially implemented ITR situation.} }\label{algo:1}
\begin{algorithmic}
\begin{scriptsize}
\Input The ITR $r\colon \mathcal{X} \mapsto \{0;1\}$, and data $(X_i^T, A_i)_{1 \leq i \leq n}$ where $X_i^{\pi^{_{s0}}}$ and $X_i^{\rho}$ are two relevant subsets of the variables contained in $X_i$. 

\Initialize the prior probabilities associated with the nodes of the tree as $g_{0,i} \gets 0.5$, and $g_{1,i} \gets 0.5$. 
\Initialize the parameters $\zeta$ of the expert $\pi^{s=0}(\cdot)$ at random e.g., $\zeta \sim \mathcal{N}(0, D)$ with $D$ a diagonal matrix.

\Compute individual contributions to $r$'s likelihood as $P_{1,i} \gets r(X_i)^{A_i} \{1-r(X_i)\}^{1-A_i}$.

\Compute individual predictions from the initiated expert network $\pi^{s=0}(\cdot)$ as $p_{0,i} \gets \expit(\zeta^TX_i^{\pi^{_{s0}}})$.

\Iterate until convergence on the parameters $\zeta$: \\
Compute individual contributions to $\pi^{s=0}$'s likelihood as $P_{0,i} \gets p_{0,i}^{A_i} \big(1-p_{0,i}\big)^{1-A_i}$. \\ 
Compute the posterior probabilities associated with the nodes of the tree as \Comment{E-step}\\
\hspace{0.8cm} $h_{0,i} \gets \frac{g_{0,i}P_{0,i}}{g_{0,i}P_{0,i} + g_{1,i}P_{1,i}} \quad$ and $\quad h_{1,i} \gets \frac{g_{1,i}P_{1,i}}{g_{0,i}P_{0,i} + g_{1,i}P_{1,i}}.$ \\
For the gating network $\rho(\cdot)$ estimate parameters $\gamma$ by solving the IRLS problem \Comment{M-step}\\
\hspace{.4cm} $\gamma \gets \argmax\limits_{\gamma} \sum_{i=1}^n h_{1,i} \ln{\big\{\expit(\gamma^TX_i^{\rho} )\big\}} + (1-h_{1,i}) \ln{\big\{1- \expit (\gamma^TX_i^{\rho}})\big\}  $ \\
For the expert network $\pi^{s=0}(\cdot)$ estimate parameters $\zeta$ by solving the IRLS problem  \\ 
\hspace{.4cm} $\zeta \gets \argmax\limits_{\zeta} \sum_{i=1}^n h_{0,i} \Big[A_i \ln{\big\{\expit(\zeta^TX_i^{\pi^{_{s0}}} )\big\}} + (1-A_i) \ln{\big\{1- \expit (\zeta^TX_i^{\pi^{_{s0}}}})\big\} \Big]  $ \\

Update the prior probabilities associated with the nodes of the tree as \\
\hspace{.8cm} $g_{1,i} \gets \expit(\gamma^TX_i^{\rho}) \quad$ and $\quad g_{0,i} \gets 1 - g_{1,i}.$ \\

Update the predictions from the expert network $\pi^{s=0}(\cdot)$ as $p_{0,i} \gets \expit(\zeta^TX^{\pi^{s=0}}_i)$.

\Endloop $\hat{\pi}^{s=0}(x)=\expit(\zeta^Tx).$
\end{scriptsize}
\end{algorithmic}
\end{algorithm}
\clearpage

\section{Simulations} \label{Sims}
\subsection{Setup}
In this section, we study the properties of the MIG, ARE and AIE estimators in the partially implemented ITR situation. To this end, we simulate data analysis in a setting where an ITR was partially implemented. We generate synthetic datasets comprising six Bernoulli, log-normally and normally distributed, correlated, covariates $X=(X_{1}, X_{2},\dots,X_{6})$ as follows. 
\begin{enumerate}[Step 1]
\item We randomly generate correlated intermediate covariates $X_{1}', X_{2}',\dots,X_{6}'$ from a multivariate gaussian distribution
    \begin{displaymath}
    \left(X_{1}', X_{2}', \ldots,  X_{6}' \right)^{T} \sim \mathcal{N}(0,\,\Sigma).
    \end{displaymath}
To generate $\Sigma$, we chose 6 eigenvalues $(\lambda_{1},\lambda_{2},\dots,\lambda_{6})=(1,1.2,1.4,1.6,1.8,2)$, and sample a random orthogonal matrix $O$ of size $6 \times 6$. The covariance matrix $\Sigma$ is obtained via
  \begin{align*}
     \Sigma &=O
     \begin{bmatrix} 
    \lambda_{1} & 0               & \dots  & 0\\
    0           & \lambda_{2}     & \ddots & \vdots\\
    \vdots      & \ddots          & \ddots & 0\\
    0           & \dots           & 0      & \lambda_{6}
    \end{bmatrix}
    O^{T}.
  \end{align*}
\item To allow for the Bernoulli or log-normal distribution of covariates, we generate $X_{1}, X_{2},\dots,X_{6}$ as follows 
$ (X_{1},X_{2}) =(\mathbbm{1}\{X_{1}'<0\},\mathbbm{1}\{X_{2}'<0\}) $,
$ (X_{3},X_{4},X_{5}) =\big(\exp(X_{3}'),\exp(X_{4}'),\exp(X_{5}')\big)$,
$X_{6} =X_{6}'$. We add $X_{0} \equiv 1$ to allow for intercepts.
\item We generate data from the covariates in this manner:
\begin{multicols}{2}
\noindent
  \begin{align*}
&S|X=x \sim \text{Bernouilli}\big( \expit(\gamma^Tx)\big), \\
&r(X)  = \mathbbm{1} \{ \delta^Tx <0 \}, \\
&A^{s=0}|X=x  \sim \text{Bernouilli}\big( \expit(\zeta^Tx)\big),  \\
&A^{s=1}  =r(X), \\
&A =  SA^{s=1} + (1-S)A^{s=0}, \\
    \end{align*} 
    \begin{align*}
&Y^{a=0}|X=x \sim \text{Bernouilli}\big( \expit(\alpha^Tx)\big), \\
&Y^{a=1}|X=x \sim \text{Bernouilli}\big( \expit(\beta^Tx)\big), \\
&Y^{s=0}  = A^{s=0}Y^{a=1} + (1-A^{s=0})Y^{a=0}, \\
&Y^{s=1}  = A^{s=1}Y^{a=1} + (1-A^{s=1})Y^{a=0}, \\
&Y =  AY^{a=1} + (1-A)Y^{a=0} 
  \end{align*}
\end{multicols}
\vspace{-1cm} with
\begin{align*}
\gamma=&(0, 0, 0, 0, 0, 0, 1)^T, & &\delta= (0.05, -0.5, 0.5, -0.5, 0.5, 0, 0)^T, \\
\alpha=&(0, -0.3, -0.05, 0.5, -0.15, -0.2, 0)^T, & & \beta= (0, -0.2, 0.05, 0.3, -0.1, -0.1, 0)^T
\end{align*}
and we vary $\zeta$.
\end{enumerate}
In scenario A, we set $\zeta=\delta$ which corresponds to a situation where treatment allocation in the absence of the ITR is different from the ITR. In scenario B, we set $\zeta=(0,0,0,0,0,0,0)$ which corresponds to a situation where treatment allocation in the absence of the ITR is random with probability $0.5$. In scenario C, we set $\zeta=-\delta$ which corresponds to a situation where treatment allocation in the absence of the ITR resembles the ITR. In each scenario we generate a target population of two million individuals from which we approximate ground truth for our estimands and drew random samples. We vary the sample size: $n=200,800,2000$. The potential outcomes as well as the variable $S$ are regarded as unobserved variables. Models for $\mu_0$ and $\mu_1$ are correctly specified with $X^{\mu}=(X_2, X_3, X_4, X_5, X_6)$ as explanatory variables. We fit the mixture of expert in equation \eqref{model:score_prop} with Algorithm \ref{algo:1}, specifying the gating network $\rho$ with $X^{\rho}=X_6$ and the expert network $\pi^{s=0}$ with variables $X^{\pi^{_{s0}}}=(X_1, X_2, X_3, X_4, X_5)$. For each scenario/sample size combination, we implement 1000 simulation iterations and 999 bootstrap replications to generate confidence intervals.

\subsection{Results}
\begin{figure}[htb!]
    \centering
    \includegraphics[width=.61\textwidth]{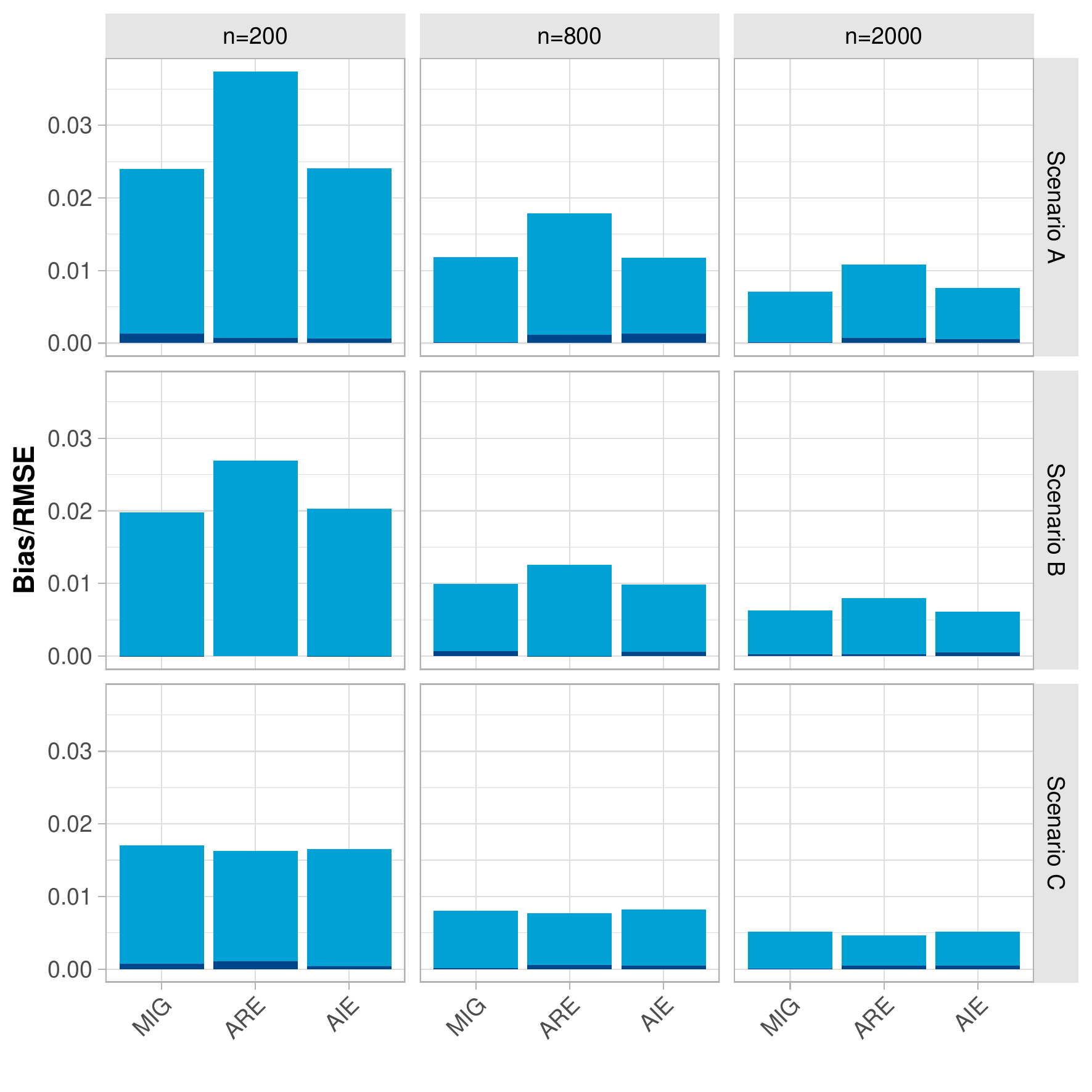}
    \caption{ Absolute bias and Root Mean Square Error (RMSE) for the MIG, ARE and AIE estimators across 1000 simulation iterations in nine scenario/sample size combinations. Absolute bias is the darker portion of each bar ; RMSE corresponds to the total bar size. The MIG, ARE and AIE estimators are from $\widehat{\Gamma}_Q(r,\rho),
    \widehat{\Delta}_{ME}(r),
    \widehat{\Lambda}_{ME}(r, \rho)o$ respectively.
    }
    \label{fig:simul}
\end{figure}

The results of our simulations are reported in Table \ref{table:simul} and Figure \ref{fig:simul}. The MIG estimator $\widehat{\Gamma}_{Q}(r,\rho)$, which does not rely on an EM procedure, exhibits its theoretical properties of unbiasness and consistency. The ARE estimator $\widehat{\Delta}_{ME}(r)$ and the AIE estimator $\widehat{\Lambda}_{ME}(r,\rho)$ which both rely on the EM procedure also appear unbiased and consistent. Their standard error is comparable to that of the MIG estimator $\widehat{\Gamma}_{Q}(r,\rho)$. Ninety five percent bootstrap confidence intervals achieve close to nominal coverage for all three estimators. 

\begin{table}[hb!]
\caption{ Simulation results for the MIG, AIE and ARE estimators under all nine scenario/sample size combinations. The MIG, ARE and AIE estimators are from $\widehat{\Gamma}_Q(r,\rho), \widehat{\Delta}_{ME}(r), \widehat{\Lambda}_{ME}(r, \rho)$ respectively. SE: standard error; RMSE: root mean squared error; CI: 95\% confidence interval. Coverage probabilities are for 95\% confidence intervals.\label{table:simul}}
\begin{center}
\begin{tabular}{lccccccccc}
	&	\multicolumn{3}{c}{Scenario A}	&		\multicolumn{3}{c}{Scenario B} & \multicolumn{3}{c}{Scenario C}		\\ \cline{2-10}
$n$	&	MIG	&	ARE	&	AIE	&	MIG	&	ARE	&	AIE	&	MIG	&	ARE	&	AIE	\\ \hline
True value	&	-0.013	&	-0.026	&	-0.013	&	-0.008	&	-0.016	&	-0.008	&	-0.004	&	-0.007	&	-0.003	\\
Bias	&		&		&		&		&		&		&		&		&		\\
\quad 200	&	-0.001	&	-0.001	&	0.001	&	0.000	&	0.000	&	0.000	&	-0.001	&	-0.001	&	0.000	\\
\quad 800	&	0.000	&	0.001	&	0.001	&	-0.001	&	0.000	&	0.001	&	0.000	&	-0.001	&	0.000	\\
\quad 2000	&	0.000	&	0.001	&	0.001	&	0.000	&	0.000	&	0.000	&	0.000	&	0.000	&	-0.001	\\
Empirical SE	&		&		&		&		&		&		&		&		&		\\
\quad 200	&	0.024	&	0.037	&	0.024	&	0.020	&	0.027	&	0.020	&	0.017	&	0.016	&	0.017	\\
\quad 800	&	0.012	&	0.018	&	0.012	&	0.010	&	0.013	&	0.010	&	0.008	&	0.008	&	0.008	\\
\quad 2000	&	0.007	&	0.011	&	0.008	&	0.006	&	0.008	&	0.006	&	0.005	&	0.005	&	0.005	\\
RMSE	&		&		&		&		&		&		&		&		&		\\
\quad 200	&	0.024	&	0.037	&	0.024	&	0.020	&	0.027	&	0.020	&	0.017	&	0.016	&	0.017	\\
\quad 800	&	0.012	&	0.018	&	0.012	&	0.010	&	0.013	&	0.010	&	0.008	&	0.008	&	0.008	\\
\quad 2000	&	0.007	&	0.011	&	0.008	&	0.006	&	0.008	&	0.006	&	0.005	&	0.005	&	0.005	\\
Coverage 	&		&		&		&		&		&		&		&		&		\\
\quad 200	&	0.964	&	0.954	&	0.937	&	0.953	&	0.976	&	0.941	&	0.936	&	0.979	&	0.944	\\
\quad 800	&	0.945	&	0.953	&	0.949	&	0.945	&	0.956	&	0.947	&	0.953	&	0.957	&	0.949	\\
\quad 2000	&	0.959	&	0.951	&	0.934	&	0.948	&	0.946	&	0.950	&	0.962	&	0.946	&	0.947	\\
CI width	&		&		&		&		&		&		&		&		&		\\
\quad 200	&	0.099	&	0.153	&	0.096	&	0.080	&	0.113	&	0.082	&	0.067	&	0.073	&	0.067	\\
\quad 800	&	0.047	&	0.071	&	0.046	&	0.039	&	0.050	&	0.039	&	0.033	&	0.031	&	0.033	\\
\quad 2000	&	0.029	&	0.044	&	0.029	&	0.024	&	0.031	&	0.024	&	0.021	&	0.019	&	0.021	\\             
\end{tabular}
\end{center}
  
\end{table}
  
\section{Applications on the MIMIC-III database}

The Multi-Parameter Intelligent Monitoring in Intensive Care III (MIMIC-III) database is a publicly available electronic health record that contain data from 53,423 patients hospitalized in intensive care at Beth Israel Deaconess Medical Center from 2001 to 2012.\citep{Johnson2016} From these, we include the 3,748 intensive care unit adult patients with severe acute kidney injury who had received either invasive mechanical ventilation or vasopressor infusion. We report the full inclusion/exclusion criteria in the Appendix F and the inclusion flow-diagram in the Appendix G. The patients we include are eligible for recommendation from both ITRs described below. For the sake of focusing on the estimators in our methodology, we handle patients with missing data by conducting a single imputation using chained equations.\citep{White2011}

In this section, we consider two example ITRs. In the first example, we evaluate a new ITR for dialysis initiation.\footnote{In this section, we use the term “dialysis” loosely to refer to all kidney support therapies suitable for acute kidney injury patients i.e., including but not limited to intermittent hemodialysis and continuous hemofiltration.} This last ITR was not available at the time of data collection and decision to initiate dialysis never followed from its implementation. In the second example, we evaluate an ITR that was partially implemented at the time of data collection. Specifically, we evaluate the impact of an ITR that recommends initiating dialysis in the most severe patients based on the Sequential Organ Failure Assessment (SOFA) score.\citep{Vincent1996}

\subsection{New ITR: dialysis initiation based on a combination of six biomarkers}
\begin{figure}[htb!]
    \centering
    \includegraphics[width=1\textwidth]{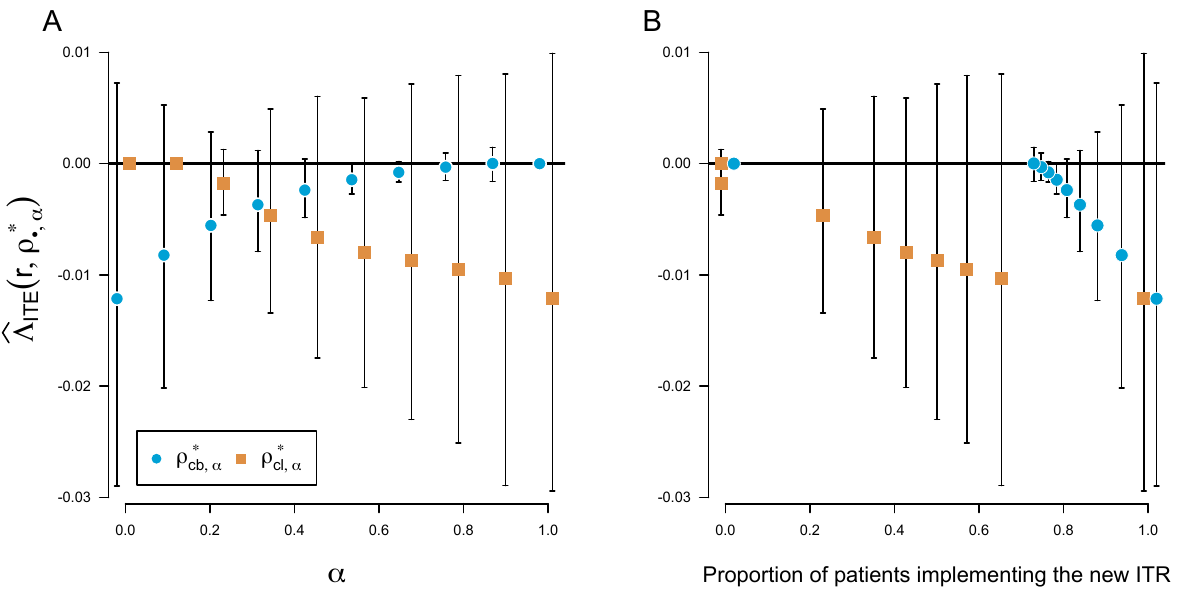}
    \caption{ Evaluation of the impact of a new ITR (i.e., dialysis initiation within 24 hours only in specific patients based on a combination of six biomarkers) on 60-day mortality using the MIMIC-III observational database. Ninety-five percent confidence intervals are from the bootstrap. Blue diamonds are for the cognitive bias scheme; orange diamonds are for the confidence level scheme. Panels A depict the AIE for different values of implementation parameter $\alpha$, Panels B depict the AIE as a function of the proportion of (future) patients implementing the new ITR: $n^{-1} \sum_{i=1}^n \rho^*_{\cdot, \alpha}(X_i)$. More negative values of the AIE indicate greater benefit from ITR implementation. Ninety-five percent confidence intervals are from the bootstrap.
    }
    \label{fig:mimic_new}
\end{figure}

Grolleau et al.\citep{Grolleau2022} have recently developed a new ITR for dialysis initiation in the intensive care unit using data from two RCTs. Briefly, this new ITR recommends initiating dialysis within 24 hours only in specific patients based on a combination of six biomarkers (SOFA score, pH, potassium, blood urea nitrogen, weight and, the prescription of immunosuppressive drug). Following the methodology detailed in section \ref{subsub:newITR}, we evaluate the impact of this new ITR on 60-day mortality. We estimate the ARE using a double robust estimator as detailed in section \ref{subsub:newimpARE}. We explore the impact of various degrees of implementation under either cognitive bias or confidence level schemes. The estimated values of AIE are given in Figure \ref{fig:mimic_new}.  Estimation of the ARE shows a trend for benefit from the implementation of the new ITR ($\widehat{\Delta}_{AIPW}(r)=$ -0.02; 95\% confidence interval [-0.06 to 0.01]). Note that the ARE estimate is not equal to estimation of the AIE under full implementation as, contrary to the ARE case, for the AIE we use a CATE model. The variables included in each model are reported in Appendix H.

\subsection{Partially implemented ITR: dialysis initiation based on SOFA scores}
\begin{figure}[htb!]
    \centering
    \includegraphics[width=.8\textwidth]{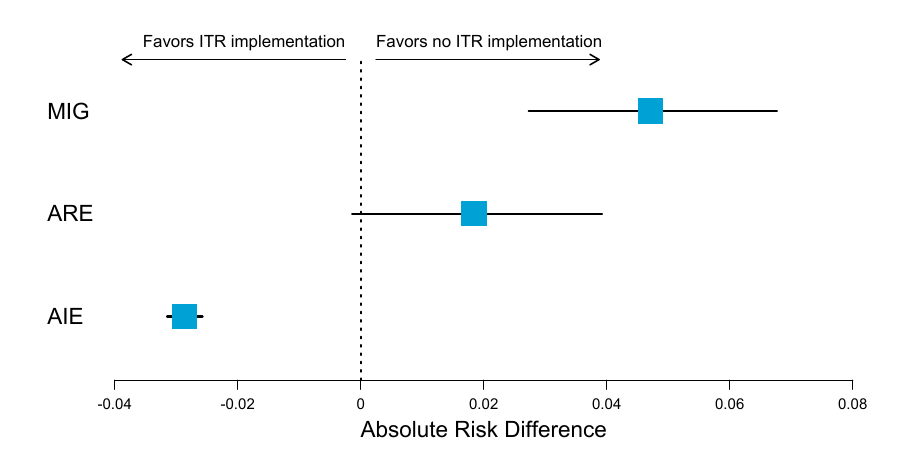}
    \caption{ Evaluation of the impact of a partially implemented ITR (i.e., dialysis initiation within 24 hours only in the patients with a SOFA score greater than 11) on 60-day mortality using the MIMIC-III observational database. Ninety-five percent confidence intervals are from the bootstrap. MIG=Maximal Implementation Gain. ARE=Average Rule Effect. AIE=Average Implementation Effect.
    }
    \label{fig:mimic_sofa}
\end{figure}
We evaluate the impact on 60-day mortality of a partially implemented, yet never evaluated, ITR that recommended initiating dialysis within 24 hours only in the patients with a Sequential Organ Failure Assessment (SOFA) score greater than 11. Following the methodology of section \ref{subsub:iplITR}, we posit models for the treatment-specific prognosis functions, the propensity score in the absence of ITR implementation (expert network) and the stochastic implementation function (gating network). Specification of propensity score in the absence of ITR implementation include the variables thought to have caused treatment initiation while specification of the stochastic implementation function include all variables thought to be associated with ITR implementation. The variables included in each model are reported in I. The estimates of the MIG, ARE and AIE are given in Figure \ref{fig:mimic_sofa}. Estimation of the MIG shows evidence of harm from further implementing the ITR ($\widehat{\Gamma}_Q(r,\rho)=$ 4.7\%; 95\% confidence interval [2.7\% to 6.8\%]). Similarly, estimation of the ARE shows a trend for harm in implementing the ITR in all patients versus in no one ($\widehat{\Delta}_Q(r)=$ 1.8\%; 95\% confidence interval [-0.1\% to 3.9\%]) indicating that the ITR may be poorly designed. However, estimation of the AIE shows that the withdrawal of the ITR would on average yield outcomes worse than in the current situation ($\widehat{\Lambda}_Q(r,\rho)=$ -2.9\%; 95\% confidence interval [-3.1\% to -2.6\%]). This suggest that even though the ITR may be poorly designed, physicians identify correctly the patients who benefit from ITR implementation. In sum, these results indicate that neither full nor null implementation of the ITR would improve patient outcomes (at the population level). Rather, either one of these changes in ITR implementation, our analysis suggest, would worsen patient outcomes (at the population level). From a policy-maker standpoint, the best thing to do under such conditions may be to develop and subsequently evaluate a new ITR.

\section{Discussion}
Our goal was to construct an ecosystem for the evaluation of ITRs that will ultimately benefit patients. We believe that the probability model and inferential approach we introduced in this paper provide actionable tools to move this agenda forward. Below, we discuss some limitations of our approach. \par
In the \emph{new ITR situation}, the exploration of the AIE relies on assuming future implementation schemes. Though sensible, the three implementation schemes we propose are subjective. Other realistic implementation schemes can be assumed and readily implemented in our methodology. \par
In the \emph{partially implemented ITR situation}, inference relies on assuming a new probabilistic model. This model is largely inspired by the Neyman-Rubin causal model. As in the original model, our model requires assuming that the effect of an ITR is mediated only by the treatment prescribed by physicians to their patients. This exclusion restriction assumption may not hold in some specific settings. In cases where the decision to implement the ITR is taken by the patients—not the physicians—, it is possible that merely seeing an ITR's recomendation affects outcomes. For instance, if an ITR recommends a patient treatment A, this patient may choose treatment B and compensate for not implementing the ITR by taking another effective treatment say C. This indirect effect of the ITR through treatment C would not be accounted for in our framework. With respect to the exchangeability assumption, our methodology relies on expert knowledge of the variables causing ITR implementation. This could include information about patients' physicians. For some research questions, the relevant variables may not be available thereby, limiting the usefulness of our approach. However, we do not believe that any statistical method can provide helpful workarounds under such conditions. Finally, as in the usual average treatment effect, one may be tempted to estimate the prognostic function $q_0$ directly, rather than estimating the propensity score $\pi$ from a mixture model. As $Y^{s=0}_i$ are not observed, this would require to posit a hierarchical mixture of experts model. Though compelling at first glance, this approach may be impractical as hierarchical mixture of experts were shown to have likelihoods with arbitrary bad local maxima yielding EM algorithms sensitive to initialization conditions.\cite{Jin2016} In contrast, for the simpler mixture considered in this paper, we were able to prove identifiability under mild technical conditions. Our simulations suggest that our mixture model's likelihood is well-behaved on finite samples so that an EM algorithm with a random initialization is successful with high probability in estimating the true value of the parameters. In fact, we expect that, for the specific mixture considered in this paper, under some conditions on $(X_i,A_i)_{1\leq i\leq n}$ and $r$, the model's likelihood is a concave function of the parameters. As it turns out to be mathematically challenging to give a satisfying statement of these conditions, we plan to dedicate future theoretical work to this question. \par
Other future directions for our work include extensive simulations with misspecified networks and non-parametric networks as well as the extension of the proposed framework to non-binary treatments and dynamic ITRs.

\section*{Acknowledgments}
We thank Dr. Viet Thi Tran and Prof. Stéphane Gaudry for their insightful comments on epidemiological and critical care applications.


\textcolor{white}{\appendix}

\clearpage


\section*{Supplementary material}

## A   Proof for the ARE formula in the new ITR situation

$$\begin{aligned}
\Delta(r) &= \mathbb{E}\left[Y^{s=1} - Y^{s=0}\right] \\
&= \mathbb{E}\left[r(X)Y^{a=1} + \{1 - r(X)\}Y^{a=0} - AY^{a=1} - (1-A)Y^{a=0}\right] \\
&= \mathbb{E}\left[\{r(X) - A\}\{Y^{a=1} - Y^{a=0}\}\right] \\
&= \mathbb{E}\left[\mathbb{E}\left[\{r(X) - A\}\{Y^{a=1} - Y^{a=0}\}|X\right]\right] \\
&= \mathbb{E}\left[\{r(X) - \pi(X)\}\tau(X)\right] && \text{(A1)}
\end{aligned}$$

where equality in (A1) uses equation (5) from assumption 2 and the fact that in the new ITR situation $A = A^{s=0}$.

## B   Proof for the AIE/MIG formulas in the new ITR situation

$$\begin{aligned}
\Lambda(r, \rho^*) &= \mathbb{E}\left[Y^* - Y^{s=0}\right] \\
&= \mathbb{E}\left[A^*Y^{a=1} + (1 - A^*)Y^{a=0} - A^{s=0}Y^{a=1} - (1 - A^{s=0})Y^{a=0}\right] \\
&= \mathbb{E}\left[\{A^* - A\}\{Y^{a=1} - Y^{a=0}\}\right] \\
&= \mathbb{E}\left[\mathbb{E}\left[\{A^* - A\}\{Y^{a=1} - Y^{a=0}\}|X\right]\right] \\
&= \mathbb{E}\left[\{\pi^*(X) - \pi(X)\}\tau(X)\right] && \text{(B1)} \\
&= \mathbb{E}\left[\rho^*(X)\{r(X) - \pi(X)\}\tau(X)\right] && \text{(B2)}
\end{aligned}$$

where equality in (B1) uses equations (5) and (6) from assumption 2 and the fact that in the new ITR situation $A = A^{s=0}$. The equality in (B2) follows from

$$\begin{aligned}
\pi^*(X) &= \mathbb{E}[A^*|X] \\
&= \mathbb{E}[S^*A^{s=1} + (1 - S^*)A^{s=0}|X] \\
&= \rho^*(X)r(X) + \{1 - \rho^*(X)\}\pi(X)
\end{aligned}$$

where the last equality uses the fact that $A^{s=1} = r(X)$ and equation (7) from assumption 2. The proof for the $\Gamma(r, \rho^*)$ formula follows a similar argument.

## C  Proof of lemma 1 in the partially implemented ITR situation

$$
\begin{aligned}
(i) \quad \pi(X) &= \mathbb{E}[A|X] \\
&= \mathbb{E}[Sr(X) + (1-S)A^{s=0}|X] \\
&= r(X)\,\mathbb{E}[S|X] + \mathbb{E}[(1-S)A^{s=0}|X] \\
&= r(X)\rho(X) + \mathbb{E}[1-S|X]\,\mathbb{E}[A^{s=0}|X] \\
&= r(X)\rho(X) + \{1-\rho(X)\}\pi^{s=0}(X)
\end{aligned} \tag{C1}
$$

Equality in (C1) relies on equation (7) from assumption 2 and the fact that in the partially implemented ITR situation $S = S^*$.

$$
\begin{aligned}
(ii) \quad q_0(X) &= \mathbb{E}[Y^{s=0}|X] \\
&= \mathbb{E}[A^{s=0}Y^{a=1} + (1-A^{s=0})Y^{a=0}|X] \\
&= \mathbb{E}[A^{s=0}Y^{a=1}|X] + \mathbb{E}[(1-A^{s=0})Y^{a=0}|X] \\
&= \mathbb{E}[A^{s=0}|X]\,\mathbb{E}[Y^{a=1}|X] + \mathbb{E}[(1-A^{s=0})|X]\,\mathbb{E}[Y^{a=0}|X] \\
&= \pi^{s=0}(X)\mu_1(X) + \{1-\pi^{s=0}(X)\}\mu_0(X)
\end{aligned} \tag{C2}
$$

Equality in (C2) relies on equation (5) from assumption 2.

## D   Identifiablity result

In this appendix, we prove an identifiability result for the model given by equation (3.14)

$$\pi(x, \theta) = \rho(x, \alpha, a)r(x) + \{1 - \rho(x, \alpha)\}\pi^{s=0}(x, \beta, b)$$

where $\theta = (\alpha, a, \beta, b) \in \mathbb{R}^{2d+2}$.

**Theorem.** *Assume that $\mathcal{X}$ is a open subset of $\mathbb{R}^d$, that $r \colon \mathcal{X} \to \{0;1\}$ is such that the sets $\{r = 0\}$ and $\{r = 1\}$ have non-empty interior and that the applications $\rho(\cdot, \alpha, a) \colon \mathcal{X} \to \mathbb{R}$ and $\pi^{s=0}(\cdot, \beta, b) \colon \mathcal{X} \to \mathbb{R}$ are of the form*

$$\rho(x, \alpha, a) = \text{expit}(\alpha^T x + a),$$
$$\pi^{s=0}(x, \beta, b) = \text{expit}(\beta^T x + b).$$

*Then the model $\pi(x, \theta)$ is identifiable.*

*Proof.* Let $\theta$ and $\theta'$ in $\mathbb{R}^{d+2}$ such that $\pi(x, \theta) = \pi(x, \theta')$. Consider the functions

$$f(x) \colon \mathbb{R}^d \to \mathbb{R}, \ x \mapsto \{(1 - \rho(x, \alpha, a))\pi^{s=0}(x, \beta, b)\}^{-1}$$
$$g(x) \colon \mathbb{R}^d \to \mathbb{R}, \ x \mapsto \{(1 - \rho(x, \alpha', a'))\pi^{s=0}(x, \beta', b')\}^{-1}$$

Since exp is an entire function, $f$ and $g$ can be written as power i.e. there exist sequences of real numbers $c_\nu$ and $c'_\nu$ such that for all $x \in \mathbb{R}^d$,

$$f(x) = \sum_{\nu \in \mathbb{N}^d} c_\nu x^\nu \qquad \text{and} \qquad g(x) = \sum_{\nu \in \mathbb{N}^d} c'_\nu x^\nu \tag{D3}$$

where $\nu = (\nu_1, \nu_2, \ldots, \nu_d)^T \in \mathbb{N}^d$, $x = (x_1, \ldots, x_d)^T \in \mathbb{R}^d$, and $x^\nu$ is a shorthand for $x_1^{\nu_1} x_2^{\nu_2} \ldots x_d^{\nu_d}$. Moreover, we have the following identities

$$\forall x \in \{r = 0\}, \ (1 - \rho(x, \alpha, a))\pi^{s=0}(x, \beta, b)$$
$$= (1 - \rho(x, \alpha', a'))\pi^{s=0}(x, \beta', b'), \tag{D4}$$

$$\forall x \in \{r = 1\}, \ \rho(x, \alpha, a) + \{1 - \rho(x, \alpha)\}\pi^{s=0}(x, \beta, b)$$
$$= \rho(x, \alpha', a') + \{1 - \rho(x, \alpha')\}\pi^{s=0}(x, \beta', b'). \tag{D5}$$

Equation (D4) implies that on $\{r = 0\}$, $f \equiv g$. Since, $f$ and $g$ are power series that coincide on $\{r = 0\}$ which has a non-empty interior, the principle of analytic continuation implies that $f \equiv g$ on $\mathbb{R}^d$. Hence, Equation (D4) holds on

$\mathbb{R}^d$ and in particular on $\{r = 1\}$. Substracting Equation (D4) from Equation (D5), we obtain that

$$\forall x \in \{r = 1\}, \ \rho(x, \alpha, a) = \rho(x, \alpha', a')$$

Since $\exp it : \mathbb{R} \to \mathbb{R}$ is a bijection, we have that for every $x \in \{r = 1\}$, $\alpha^T x + a = \alpha'^T x + a'$. Since $\{r = 1\}$ has a non-empty interior, we get $(\alpha, a) = (\alpha', a')$. This last equality together with Equation (D4) implies that for every $x \in \mathbb{R}^d \ \pi^{s=0}(x, \beta, b) = \pi^{s=0}(x, \beta', b')$. It follows that $(\beta, b) = (\beta', b')$. Hence the model $\pi(x, \theta)$ is identifiable. $\qquad\square$

# E    Derivation of an EM algorithm for the mixture of experts presented in section 3.2.2

Here, we follow the structure of the exposition of Jordan and Jacobs (1994). For simplicity, we start assuming that the true models for the expert $\pi^{s=0}$ and the gating network $\rho$ are parametric models. Under this assumption, we characterize the probability of generating $A_i$ from $X_i$ separately for the case where $S_i = 1$ and the case where $S_i = 0$. When $S_i = 1$, this probability is either 0 or 1 that is,

$$
\begin{aligned}
P_1(A_i|X_i) &= \mathbb{1}\{A_i = r(X_i)\} \\
&= r(X_i)A_i + \{1 - r(X_i)\}(1 - A_i) \\
&= r(X_i)^{A_i}\{1 - r(X_i)\}^{1-A_i}.
\end{aligned}
$$

When $S_i = 0$, the probability is

$$
P_0(A_i|X_i, \zeta^0) = \pi^{s=0}(X_i; \zeta^0)^{A_i}\left(1 - \pi^{s=0}(X_i; \zeta^0)\right)^{1-A_i}
$$

where $\zeta^0$ denotes the true value of the parameter for the expert $\pi^{s=0}$. The total probability of generating $A_i$ from $X_i$ is thus

$$
P(A_i|X_i, \theta^0) = \rho(X_i; \gamma^0)P_1(A_i|X_i) + \{1 - \rho(X_i; \gamma^0)\}P_0(A_i|X_i, \zeta^0) \qquad \text{(E1)}
$$

where we denote $\gamma^0$ the true value of the parameter for the gating network $\rho$ and we write $\theta^0 = (\gamma^0, \zeta^0)^T$. We utilize (E1) without the superscripts to refer to the probability model defined by the mixture, irrespective of any reference to a "true" model. Given a data set $\mathcal{D} = (X_i^T, A_i)_{1 \leq i \leq n}$, the log-likelihood is obtained by taking the log of the product of $n$ densities of the form of (E1), which yields the following log-likelihood:

$$
l(\theta; \mathcal{D}) = \sum_{i=1}^{n} \ln\left\{\rho(X_i; \gamma)P_1(A_i|X_i) + \{1 - \rho(X_i; \gamma)\}P_0(A_i|X_i, \zeta)\right\}. \qquad \text{(E2)}
$$

We refer to (E2) as the "incomplete-data log-likelihood" as it involves no reference to the latent random variable $S_i$. If $S_i$ were known, then the maximum likelihood problem would decouple into a separate set of two regression problems: one for the gating network $\rho$ and another for the unknown expert $\pi^{s=0}$. These problems would be solved independently of each other, yielding a rapid one-pass learning algorithm. Of course, $S_i$ are not known, but we can specify a probability model that links them to the observable data. To this end, we posit

the following probability model written in terms of the $S_i$:

$$P(A_i, S_i | X_i, \theta) = \left\{ \rho(X_i; \gamma) P_1(A_i | X_i) \right\}^{S_i} \left[ \left\{ 1 - \rho(X_i; \gamma) \right\} P_0(A_i | X_i, \zeta) \right]^{1-S_i}.$$

Taking the logarithm of this probability model yields the following "complete-data log-likelihood":

$$l_c(\theta; \mathcal{C}) = \sum_{i=1}^{n} S_i \ln \left\{ \rho(X_i; \gamma) P_1(A_i | X_i) \right\} + (1 - S_i) \ln \left\{ \left( 1 - \rho(X_i; \gamma) \right) P_0(A_i | X_i, \zeta) \right\}$$
(E3)

where $\mathcal{C} = (X_i^T, A_i, S_i)_{1 \leq i \leq n}$ denotes a "complete data" set. Note the relationship of the complete-data log-likelihood in equation (E3) to the incomplete-data log-likelihood in equation (E2). The use of the latent variables $S_i$ allowed the logarithm to be broken into a sum of logarithms, substantially simplifying the maximization problem. However, notice that the complete-data likelihood is a random variable, because the variables $S_i$ are in fact unknown. Taking expectation with respect to the value of the parameters at the $p$-th iteration denoted $\theta^{(p)}$, we can conduct the E step of the EM algorithm to obtain the deterministic (i.e., free of the random variables $S_i$) function $Q$:

$$\begin{aligned}
Q(\theta, \theta^{(p)}) &\stackrel{\text{def}}{=} \mathbb{E}[l_c(\theta; \mathcal{C}) | \mathcal{D}] \\
&= \sum_{i=1}^{n} h_{1,i} \ln \rho(X_i; \gamma) + (1 - h_{1,i}) \ln \left( 1 - \rho(X_i; \gamma) \right) \\
&\quad + \sum_{i=1}^{n} (1 - h_{1,i}) \left[ A_i \ln \left\{ \pi^{s=0}(X_i; \zeta) \right\} + (1 - A_i) \ln \left\{ 1 - \pi^{s=0}(X_i; \zeta) \right\} \right] \\
&\quad + \sum_{i=1}^{n} h_{1,i} \ln P_1(A_i | X_i)
\end{aligned}$$
(E4)

where we denote $h_{1,i} = \frac{P_1(A_i | X_i) \rho(X_i; \gamma^{(p)})}{\rho(X_i; \gamma^{(p)}) P_1(A_i | X_i) + (1 - \rho(X_i; \gamma^{(p)})) P_0(A_i | X_i, \zeta^{(p)})}$ and we have used the fact that

$$\begin{aligned}
\mathbb{E}[S_i | \mathcal{D}] &= \mathbb{P}(S_i = 1 | A_i, X_i, \theta^{(p)}) \\
&= \frac{\mathbb{P}(A_i | S_i = 1, X_i, \theta^{(p)}) \mathbb{P}(S_i = 1 | X_i, \theta^{(p)})}{\mathbb{P}(A_i | X_i, \theta^{(p)})} \\
&= \frac{\mathbb{P}(A_i | S_i = 1, X_i, \theta^{(p)}) \mathbb{P}(S_i = 1 | X_i, \theta^{(p)})}{\mathbb{P}(S_i = 1 | X_i, \theta^{(p)}) \mathbb{P}(A_i | S_i = 1, X_i, \theta^{(p)}) + \mathbb{P}(S_i = 0 | X_i, \theta^{(p)}) \mathbb{P}(A_i | S_i = 0, X_i, \theta^{(p)})} \\
&= \frac{P_1(A_i | X_i) \rho(X_i; \gamma^{(p)})}{\rho(X_i; \gamma^{(p)}) P_1(A_i | X_i) + (1 - \rho(X_i; \gamma^{(p)})) P_0(A_i | X_i, \zeta^{(p)})} \\
&= h_{1,i}.
\end{aligned}$$

The M step requires maximizing $Q(\theta, \theta^{(p)})$ with respect to the gating network parameters $\gamma$ and the expert parameters $\zeta$. Examining equation (E4), we see that the parameters $\gamma$ and $\zeta$ influence the $Q$ function only through the first and second summation terms. Thus the M step reduces to the following separate maximization problems:

$$\gamma^{(p+1)} = \arg\max_{\gamma} \sum_{i=1}^{n} h_{1,i} \ln \rho(X_i; \gamma) + (1 - h_{1,i}) \ln \left( 1 - \rho(X_i; \gamma) \right) \qquad \text{(E5)}$$

$$\zeta^{(p+1)} = \arg\max_{\zeta} \sum_{i=1}^{n} (1 - h_{1,i}) \left[ A_i \ln \left\{ \pi^{s=0}(X_i; \zeta) \right\} + (1 - A_i) \ln \left\{ 1 - \pi^{s=0}(X_i; \zeta) \right\} \right] \qquad \text{(E6)}$$

Each of these maximization problems is itself a maximum likelihood problem. Equation (E5) is a maximum likelihood problem with observations $(X_i^T, h_{1,i})_{1 \leq i \leq n}$. Equation (E6) is a weighted maximum likelihood problem with observations $(X_i^T, A_i)_{1 \leq i \leq n}$ and observation weights $1 - h_{1,i}$. Conducting the M-step maximizes the value of Q, the expectation of the complete-data log-likelihood. (Dempster *and others*, 1977) proved than an increase in $Q$ implies an increase in the incomplete-data log-likelihood and so we have

$$l(\theta^{(p+1)}; \mathcal{D}) \geq l(\theta^{(p)}; \mathcal{D}).$$

For clarity and ease of implementation, in Algorithm 1, we assume the Bernoulli distribution for both the expert and gating networks models and thus write $\pi^{s=0}(x; \zeta) = \text{expit}(\zeta^T x)$ and $\rho(x; \gamma) = \text{expit}(\gamma^T x)$. Note that viewing equations (E5) and (E6) as supervised learning problems with relevant observations and weights, nonparametric models such as boosting, random forests, or deep neural networks may also be used for estimating $\rho(\cdot)$ and $\pi^{s=0}(\cdot)$. That option is available, although we lose the EM proof of convergence (cf. Jordan and Xu (1995)). In Algorithm 1, we simplify notations as follows: $P_{1,i}$ and $P_{0,i}$ stand for $P_1(A_i|X_i)$ and $P_0(A_i|X_i, \zeta^{(p)})$ respectively; $g_{1,i}$ and $g_{0,i}$ stand for $\rho(X_i; \gamma^{(p)})$ and $1 - \rho(X_i; \gamma^{(p)})$ respectively; and we note $1 - h_{1,i}$ as $h_{0,i}$.

## F   Inclusion and exclusion criteria from the MIMIC-III analysis

Inclusion criteria were (all needed be fulfilled):

- Admission to an intensive care unit

- Age greater than 18 years on the day of intensive care admission

- Evidence of severe acute kidney injury (stage 3 in the Kidney Disease Improving Global Outcomes classification) during the stay in intensive care

- Initiation of either mechanical ventilation or intravenous vasopressors during the stay in intensive care, prior to severe acute kidney injury

Exclusion criteria were (all needed be absent):

- End-stage renal kidney disease at intensive care admission

- Renal replacement therapy initiated prior to severe acute kidney injury

- Patients included in the study for an earlier episode of severe acute kidney injury in intensive care

- Patients expected to die within three days

## G    Patient inclusion diagram from the MIMIC-III database

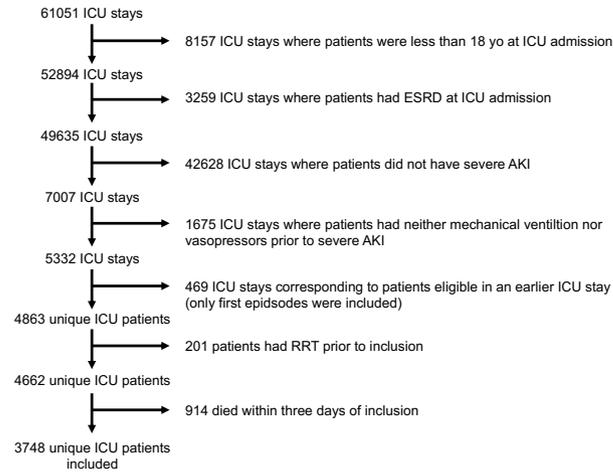

Figure G1:   Patient inclusion diagram from the MIMIC-III database. ICU=Intensive Care Unit.    ESRD=End-Stage Kidney Renal Disease. AKI=Acute Kidney Injury. RRT=Renal Replacement Therapy.

## H Model specification in the MIMIC-III analysis for the new ITR: dialysis initiation based on a combination of six biomarkers

For this analysis, we specify the models as follows.

- Propensity score model
  $\hat{\pi}(X^{\pi}) = \mathbb{P}(\text{Dialysis\_initiation\_within\_24h}|\text{Patient\_characteristics})$
  $\sim \text{Age+Weight+BUN+pH+Potassium+SOFA+Immunosuppressive\_drug}$

- Prognosis model
  $\hat{\mathbb{E}}[Y|X, A] = \mathbb{P}(\text{Death\_at\_day\_60}|\text{Patient\_characteristics, Dialysis\_initiation\_within\_24h})$
  $\sim \text{Age} + \text{Weight} + \text{BUN} + \text{pH} + \text{Potassium} + \text{SOFA}$
  $+\text{Dialysis\_initiation\_within\_24h} \times \Big[\text{Age+Weight+BUN+pH+Potassium+}$
  $\text{SOFA}\Big]$

- ITE model
  $\hat{\tau}(X) = \hat{\mathbb{E}}[Y|X, A = 1] - \hat{\mathbb{E}}[Y|X, A = 0]$

- Prognostic function under the ITR
  $\hat{q}_1(X) = r(X)\hat{\mathbb{E}}[Y|X, A = 1] + \{1 - r(X)\}\hat{\mathbb{E}}[Y|X, A = 0]$

BUN=Blood Urea Nitrogen. SOFA=Sequential Organ Failure Assessment. ITE=Individualized Treatment Effect. ITR=Individualized Treatment Rule.

# I    Model specification in the MIMIC-III analysis for a partially implemented ITR: dialysis initiation based on SOFA scores

For this analysis, we specify the models as follows.

- Treatment-specific prognosis functions

  $\hat{\mu}_0(X^\mu) = \mathbb{P}(\text{Death\_at\_day\_60}|\text{Patient\_characteristics})$
  $\sim \text{Age} + \text{SOFA}$

  $\hat{\mu}_1(X^\mu) = \mathbb{P}(\text{Death\_at\_day\_60}|\text{Patient\_characteristics})$
  $\sim \text{Age} + \text{SOFA}$

- Propensity score in the absence of ITR implementation (expert network)

  $\hat{\pi}^{s=0}(X^{\pi^{s0}}) = \mathbb{P}(\text{Dialysis\_initiation\_within\_24h}|\text{Patient\_characteristics}, S = 0)$

  $\sim \text{Age} + \text{Weight} + \text{BUN} + \text{pH} + \text{Potassium} + \text{SOFA}$

- Stochastic implementation function (gating network)

  $\hat{\rho}(X^\rho) \sim \text{Age} + \text{BUN} + \text{pH} + \text{Potassium}$

BUN=Blood Urea Nitrogen. SOFA=Sequential Organ Failure Assessment.



\end{document}